\documentclass[twocolumn,aps,superscriptaddress]{revtex4-1}
\usepackage{palatino}
\usepackage[latin1]{inputenc}
\usepackage{epsf}
\usepackage{amsmath,amssymb}
\usepackage{latexsym}
\usepackage{calc}
\usepackage{color}
\usepackage{graphicx}
\usepackage{hyperref}
\newcommand{\ben}{\begin{equation}}
\newcommand{\een}{\end{equation}}
\newcommand{\bea}{\begin{eqnarray}}
\newcommand{\eea}{\end{eqnarray}}

\def\sss{\scriptscriptstyle\rm}

\def\1s{_{1,\sss S}}
\def\2s{_{2,\sss S}}
\def\x{_{\sss X}}
\def\c{_{\sss C}}
\def\s{_{\sss S}}
\def\xc{_{\sss XC}}

\def\Hxc{_{\sss HXC}}

\def\H{_{\sss H}}

\def\ext{_{\rm ext}}

\def\br{{\bf r}}

\def\bj{{\bf j}}

\begin{document}
\title{Exact time-dependent density functional theory for non-perturbative dynamics of helium atom}
\author{Davood Dar}
\affiliation{Department of Physics, Rutgers University, Newark 07102, New Jersey USA}
\author{Lionel Lacombe}
\affiliation{Department of Physics, Rutgers University, Newark 07102, New Jersey USA}
\author{Johannes Feist}
\affiliation{Departamento de F\'isica Te\'orica de la Materia Condensada and Condensed Matter Physics Center (IFIMAC), Universidad Aut\'onoma de Madrid, Madrid, Spain}
\author{Neepa T. Maitra}
\affiliation{Department of Physics, Rutgers University, Newark 07102, New Jersey USA}
\email{neepa.maitra@rutgers.edu}
\date{\today}
\begin{abstract}
By inverting the time-dependent Kohn-Sham equation for a numerically exact dynamics of the helium atom, we show that the dynamical step and peak features of the exact correlation potential found previously in one-dimensional models persist for real three-dimensional systems. We demonstrate that the Kohn-Sham and true current-densities differ by a rotational component. The results have direct implications for approximate TDDFT calculations of atoms and molecules in strong fields, emphasizing the need to go beyond the adiabatic approximation, and highlighting caution in quantitative use of the Kohn-Sham current.
\end{abstract}

\maketitle
\section{Introduction}
For simulating dynamics of electrons in non-perturbative fields, time-dependent density functional theory~\cite{RG84,M16,TDDFTbook2012,Carstenbook} (TDDFT) has emerged as a key approach, due to its favorable system-size scaling. 
In theory, TDDFT is an exact formulation of  quantum mechanics  which provides a computationally tractable approach for tackling calculations involving many-body problems in external time-dependent fields. Mapping to a non-interacting system, the Kohn-Sham (KS) system, that exactly reproduces the one-body density allows the computation of much larger systems than with traditional wavefunction methods, with no restriction on the strength of the applied fields nor on how far the system is driven from equilibrium, see Refs.~\cite{DAGBSC17,S21,GCPRR20, SEDS21,FBMHH21,YYK17} for examples in a range of recent applications.

TDDFT does not, however,  come without its own difficulties; in particular, the exchange-correlation (xc) potential in which the many-body effects are ``hidden"  needs to be approximated  as a functional of density, and the exact xc potential depends on the density in a spatially- and time- non-local way. This dependence is neglected in adiabatic approximations used in calculations today, where the instantaneous density is inserted into  a ground-state xc approximation. 
 A crucial question is how well these approximations  accurately capture the true dynamics. The lack of memory-dependence is believed to be responsible for errors in its predictions  e.g. Refs.~\cite{RN11,RN12,HTPI14,WU08,GDRS17,BMVPS18,KVPRC20,QSAC17,GWWZ14,HKLK09,DLYU17,YYCC21}, including sometimes qualitative failures.
Still, the approximations are often accurate enough to be useful, and some characterization of when to expect the adiabatic approximation to work well has also been done~\cite{LM20b}.
Studies of the exact xc potential have been made, to compare against  approximate potentials, and also to study the impact of its features on the resulting dynamics. Such studies require two ingredients: first, an exact calculation of dynamics of interacting electrons, and second, an inversion of the TDKS equations to find the exact potential. Because of challenges in obtaining these ingredients, the studies have so far been limited to model systems~\cite{LM20b,LFSEM14,SLWM17,EFRM12,FERM13,LM18,LSWM18,RG12,FNRM16,FLNM18,HRG16,HRCLG13,AV05,RNL13,EM12,TGK08,U06} involving either one-dimension (1D) and/or two electrons, or involving only small perturbations away from the ground-state. 

In this work, the exact time-dependent KS (TDKS) potential is found for the first time, for a real three-dimensional (3D) multi-electron atom in the non-perturbative regime. We find dynamical step and peak features that have a non-local-in-space and -in-time dependence on the density. 
The results have direct implications for TDDFT calculations of atoms and molecules driven far from their ground-state, as these features are missing in adiabatic approximations. They justify the relevance of the previous 1D studies, where similar dynamical step and peak features are found  in the correlation potential. 
Moreover, the example explicitly demonstrates that the KS current-density  differs from the true current-density by a rotational component. Although this has been recognized before to be theoretically possible~\cite{AV05,MBAG02,SK16b,TK09b,GM12}, not only is this difference neglected in applications today where typically the current calculated from the KS orbitals is assumed to represent the true current~\cite{AHC18,TNY19},  but the difference has not been demonstrated for systems beyond the linear response regime.

\section{Dynamics in the Helium Atom}
The system we study is the field-free evolution of a superposition state of the Helium atom, as might be reached, for example, by applying a field that is turned off after some time. The lowest few eigenstates of this atom were found using the time-dependent close-coupling method, making a partial wave expansion in coupled spherical harmonics, and using the finite element discrete variable representation to discretize the radial degrees of freedom~\cite{Feistthesis,Feist08}. We consider here linear superpositions of the singlet ground state $1^1S_0$, denoted $\Psi_0$, and  singlet first excited state $2^1P_1$ that has angular quantum numbers $L=1$ and $M=0$, denoted here $\Psi_1$, so the exact time-evolution of the two-electron state is
\ben
    \vert\Psi(t)\rangle=\frac{1}{\sqrt{1+|a|^2}}\left(\vert \Psi_0 \rangle+a e^{-i\omega t}\vert \Psi_1\rangle\right)
    \label{eq:intstate}
\een 
where $\omega=E_{2^1P}- E_{1^1S} = 0.77980$ in atomic units (a.u.),  is the frequency with which the system oscillates.The parameter $a$ gives the relative fraction of the excited state,for example $a =1$ in the case of a 50:50 superposition. 
We aim then to find the time-dependent KS potential which reproduces the exact density of the interacting state Eq.~(\ref{eq:intstate}):
\ben
n(\br, t) = \frac{1}{1+\vert a\vert^2}\left(n_{0}(\br) + \vert a\vert^2 n_{1} (\br) + 2a n_{01}(\br)\cos(\omega t) \right)
\label{eq:dens}
\een
where $n_{q}(\br)  = 2\int \vert \Psi_q(\br, \br_2)\vert^2 d^3\br_2, q = 0, 1 $ and $n_{01}(\br) = 2\int \Psi_0(\br, \br_2) \Psi_1(\br, \br_2) d^3\br_2$ .

We note here that the results we find for the xc potential apply to far more general dynamical situations than the field-free superposition state dynamics: due to an exact condition~\cite{MBW02}, the xc potential applies to \textit{any} situation where the instantaneous interacting state is given by Eq.~(\ref{eq:intstate}) at some time $t$, and the KS state is a Slater determinant (see Appendix A). 

Now, the TDDFT xc potential depends on the choice of the initial KS state~\cite{L99,RG84,EM12}; the 1-1 density-potential mapping holds only for a given initial state, which endows $v\xc(\br,t)$ with a functional dependence on both the true and KS states, $v\xc[n; \Psi(0), \Phi(0)](\br, t)$. In principle, one can begin in any initial KS state that reproduces the density of the initial interacting state and its first time-derivative; the structure of the exact xc potential has a strong dependence on this choice~\cite{EM12,FNRM16,FLNM18,LM18,SLWM17}. The choice we make here is a Slater determinant: this is the natural choice if the state Eq.~(\ref{eq:intstate}) is reached from applying an external field to a ground-state and then turning the field off.  The Slater determinant is the natural choice in most physical situations, because they begin in the ground-state (see also discussion in Supplementary Material). One would use ground-state DFT to find the initial KS orbitals, and by the ground-state theorems, this is a Slater determinant. Since the KS evolution involves a one-body Hamiltonian, the state remains a single Slater determinant. For our two-electron  spin-singlet system, this means that  we always have a single spatial KS orbital that is doubly-occupied, and must have the form:
\begin{equation}\label{eqn:phi}
\varphi(\br,t) = \sqrt{n(\br,t)/2} e^{i\alpha(\br,t)}
\end{equation}
to reproduce the exact interacting density of Eq.~\ref{eq:dens} with the phase~$\alpha$ related to the current~$\bj$ through the equation of continuity, 
\begin{equation}\label{eq:continuity}
   \nabla\cdot \bj =   \nabla\cdot\left(n(\br,t)\nabla \alpha(\br,t)\right)= - \frac{\partial}{\partial {t}}n(\br,t)
   \end{equation}
Inverting the TDKS equation yields the exact KS potential:
\begin{equation}\label{vs}
    v_{s}(\br,t)=\frac{\nabla^2\sqrt{n(\br,t)}}{2\sqrt{n(\br,t)}}-\frac{\vert\nabla\alpha(\br,t)\vert^{2}}{2}-\frac{\partial\alpha(\br,t)}{\partial t}
\end{equation}
The exact xc potential is then obtained from 
\ben
v\xc(\br, t) = v\s(\br,t) - v\H(\br,t) - v\ext(\br,t)
\label{eq:vxc}
\een
with the Hartree potential~$v\H(\br,t) = \int \frac{n(\br',t)}{\vert \br -\br'\vert} d^3\br'$ and external potential~$v\ext(\br,t) = -2/\vert\br\vert$.  Further, one can isolate the correlation component by noting that for our choice of KS state, $v\x(\br,t) = -v\H(\br,t)/2$. 

Thus, finding the exact xc  potential reduces to solving Eq.~(\ref{eq:continuity}) for $\alpha(\br, t)$. We note here that for a different choice of initial KS state, e.g. using a two-configuration state that is more similar to that of the actual interacting state, the inversion to find $v\xc$ involves an iterative numerical procedure~\cite{RL11,NRL13,RPL15}; some examples for the 1D analog of the dynamics here can be found in Refs.~\cite{FNRM16,FLNM18,LM18}. This could be a more natural state to begin the KS calculation in some situations, e.g.  if the state was  prepared in such a superposition at the initial time, however it is inaccessible in a KS evolution that begins in the ground state, as discussed earlier. The importance of judiciously choosing the KS initial state when using an adiabatic approximation has been realized and exploited in strong-field charge-migration simulations~\cite{BHMALGSL18,CL20,FBMHH21}.

Eq.~(\ref{eq:continuity}) has the form of a Sturm-Liouville equation, which has a unique solution for $\alpha(\br, t)$ for a given boundary condition. Thanks to the azimuthal symmetry of our density ( $M=0$ at all times), we need solve this in effectively two dimensions. 
We construct an explicit matrix representation of the operator $\nabla\cdot n(\br,t)\nabla$ in spherical coordinates using the fourth order finite difference scheme subject to the following boundary conditions:
\begin{equation}\label{boundary conditions}
\alpha(\br\to\infty,t)=0  \;\;\;\;\text{and}\;\;\;\;
\frac{\partial}{\partial{\theta}}\alpha(\br,t)\vert_{\theta=\pi,0}=0\,.
\end{equation}
Choosing this boundary condition at $t=0$ yields $\alpha(\br,0) = 0$ since initially the current is zero, and  fixes our initial state as $\phi(\br,0) = \sqrt{n(\br,0)/2}$. The Runge-Gross theorem then ensures that there is a unique $v\xc(\br,t)$ that reproduces the exact $n(\br, t)$ and yields a unique $\alpha(\br,t)$ at later times ~\footnote{It is not {\it a priori} obvious that the unique solution to Eq.~(\ref{eq:continuity})  with the boundary-condition Eq.~(\ref{boundary conditions}) applied with time $t$ as a parameter is compatible with the TDKS evolution, but as the results evolve smoothly in time, it is.}.

Subject to the boundary conditions Eq.~\ref{boundary conditions}, the numerical inversion of the matrix operator $\nabla\cdot n(\br,t)\nabla$  results in the solution of Eq.~(\ref{eq:continuity}) for $\alpha(\br,t)$ which in turn when inserted into  Eq.~(\ref{vs}) yields the KS potential,  $v\s(\br,t)$ (some details in Appendix B). 
\section{Results}
Several symmetry features of the dynamics of our system simplify the analysis. The azimuthal symmetry mentioned earlier together with the 
fact that the chosen superposition is one of an $L=0$ and $L=1$ state, mean that the density, current, and potentials in the lower half-plane ($\pi/2 <\theta <\pi)$ exactly follow those in the upper half-plane ($0 <\theta <\pi/2)$ a half-cycle out of phase, $O(r, \pi-\theta, t) = O(r, \theta, t+T/2)$. (See also movies of the density, current-density, and correlation potentials in Supplementary Material). Further the simple form of the superposition means that $O(\br, T-t ) = O(\br, t)$.
Thus we  show time-snapshots only over a half cycle in the lower octant. 

\begin{figure}[h]
\includegraphics[width=0.5\textwidth]{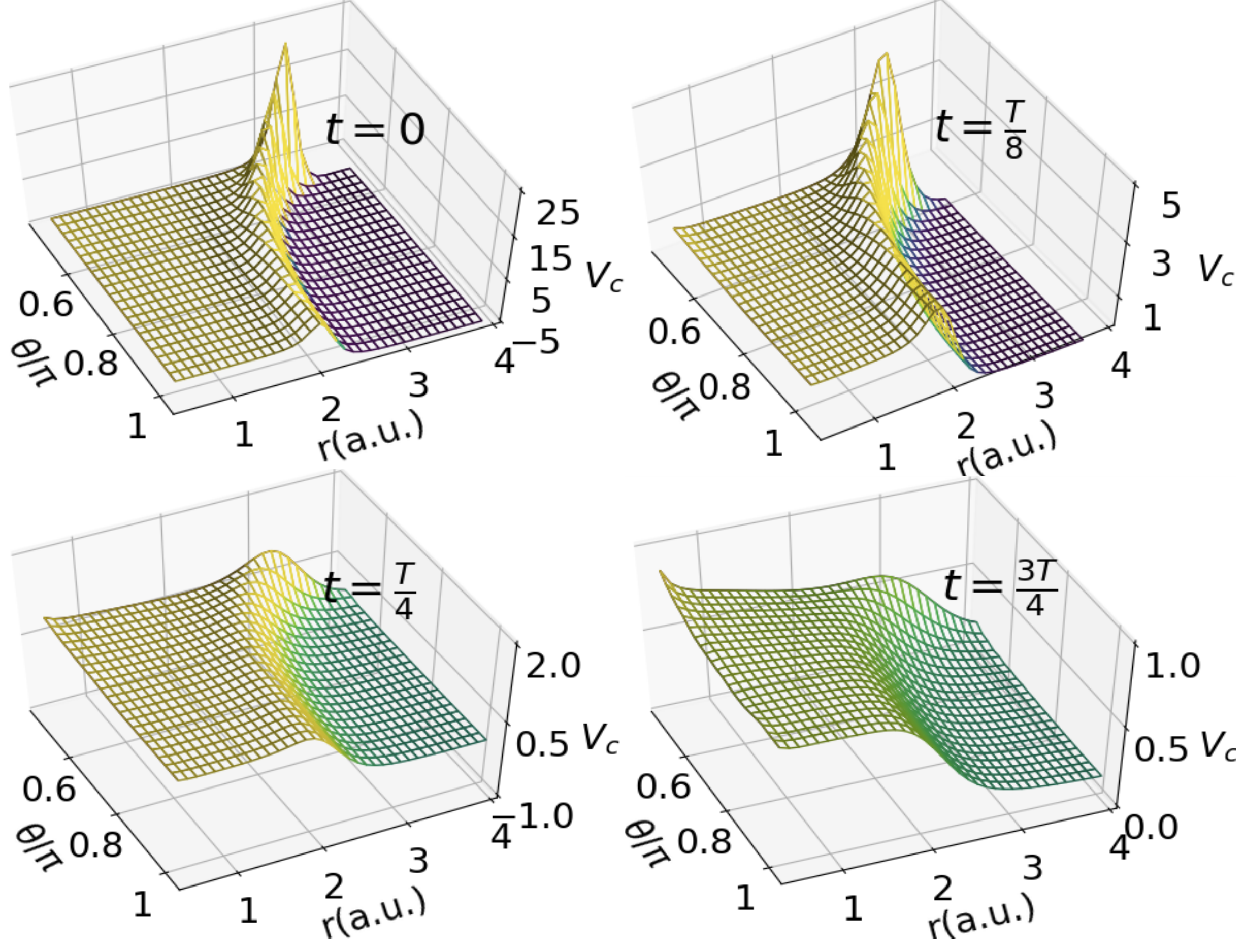} 
\caption{Correlation potential, $v\c(\br,t)$ at $t=0$ for the 50:50 superposition case ($\alpha = 1$) in the range $\pi/2 <\theta <\pi$ at times $t = 0, T/8, T/4, 3T/4$. }
\label{3Dvc}
\end{figure}

\begin{figure}[h]
     \includegraphics[width=0.5\textwidth]{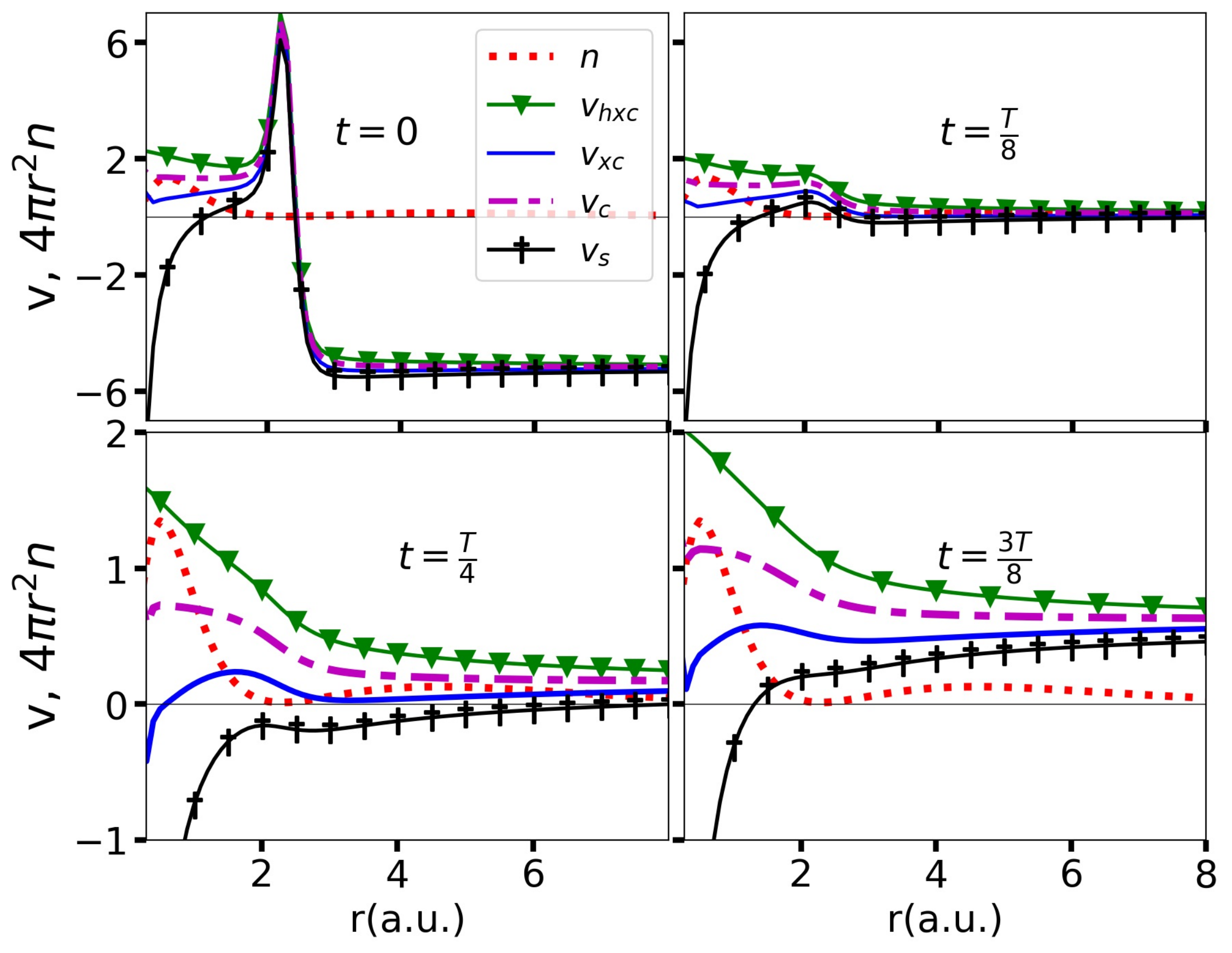} 
    \caption{Snapshots of  potentials $v\s, v\Hxc, v\xc, v\c$ and density $4\pi r^2 n$ at $\theta = 3\pi/4$, at times $t = 0, T/8, T/4, 3T/8$.}
    \label{figcpts}
\end{figure}

Figure~\ref{3Dvc}  shows  snapshots of the correlation potential indicated by fractions of the period of oscillation, $T = 2\pi/\omega = 8.057$ a.u.  One immediately notices the unmistakable presence of the step and peak features in the exact correlation potential that have been shown to arise in many 1D model systems~\cite{LM18,LM20b,LFSEM14,SLWM17,EFRM12,FERM13,LSWM18,RG12,FNRM16,FLNM18,HRG16,HRCLG13}. 
The step and peak feature is initially most prominent in the region swept by $\pi/2<\theta<\pi$, and then decreases in magnitude, gliding out of this region and appearing on the other side of $\theta=\pi/2$ at $t = T/2$. As in the 1D case, this time-dependent step 
has  a spatially nonlocal and non-adiabatic dependence on the density and is completely unaccounted for in the adiabatic approximations: They are missing even in the exact adiabatic approximation, i.e. evaluating the exact ground-state xc potential on the instantaneous density~\cite{EFRM12,FERM13,LFSEM14,SLWM17,LSWM18}.
These features often dominate the KS potential  (see Fig.~\ref{figcpts}) and have been shown to be responsible for various errors in simulations using adiabatic approximations  in 1D e.g.~\cite{SLWM17,LSWM18}. Here we find they persist just as vigorously, with the same order of magnitude, in real 3D atoms driven far from their ground-state. This justifies the relevance for real systems of the conclusions drawn from the 1D studies, and shows that such strong correlation effects are not a consequence of reduced dimensionality, as might have been assumed from ground-state systems~\cite{Brus2014}.
These dynamical steps are distinct to those arising from fractional charges~\cite{TGK08,KHG21,M17},  or in response situations~\cite{GSGB99}, as in the 1D case, and we expect they generically appear when a system is driven far from its ground-state.

 For the dynamics of this particular superposition state, at any instant of time, the correlation potential asymptotes to the same (time-dependent) value in every direction in the lower half plane, while asymptoting to a different value in the upper half plane (recall $O(\br, T - t) = O(\br, t)$). At $\theta = \pi/2$ there is a step and peak in $v\c$ in the $\theta-$direction which gives a force that ensures KS currents, like the true currents, do not cross the $xy$-plane. This complex structure makes the inversion numerically unreliable right at $\theta = \pi/2$. Along the $\theta = \pi/2$ plane, the density of the $P$ state vanishes, and the large change in the potential may be somewhat reminiscent of the abnormal divergent behavior along the HOMO nodal plane found in the ground-state potential~\cite{GGB16}. Unlike the ground-state case, however, our density does decay differently along that plane than in other directions, and moreover cannot be captured by any adiabatic approximation.

Decomposing the terms in Eq.~(\ref{vs}), we find that the peak tends to arise from the second term while the third term results in the step. Because the
KS current $\bj\s = n\nabla\alpha$, the second term and the peak are related to the local velocity $\bj\s/n$, while the step when a cut is taken across a fixed $\theta$ is related to the radial integral of the local acceleration, $\dot\alpha(r,t) = \int^r \nabla\dot\alpha(r',t).\hat{r}' dr'$.

We note that the appearance of such dominating steps in the correlation potential is fundamentally linked to the 
 difference in configurations of the interacting 
and KS states: 
the interacting system is a superposition of a ground and excited state, quite distinct
from the KS Slater determinant structure. Tuning down the electron-interaction dampens the peak structure but the step remains.

Fig.~\ref{figcpts} shows the components of the exact KS potential. We observe that in the central region where most of the density is localized, the force from the correlation potential is much smaller than that from the exchange and Hartree terms. It is in fact of similar magnitude to that in the ground-state~\cite{UG94}. 
Near the density-minimum, where the excited state begins to dominate over the ground-state, the correlation potential rises, and then falls before leveling out. In this region, the slopes are such that the step appears to be keeping different parts of the density separate, while the peak corrects for dynamical Coulombic electron-interaction effects. The lack of these features in the adiabatic approximations suggests that the resulting densities will not be as structured, and will tend to underestimate oscillation amplitudes in the dipole moment (as seen in the 1D case~\cite{FLNM18}). 

\begin{figure}[h]
\includegraphics[width=0.5\textwidth]{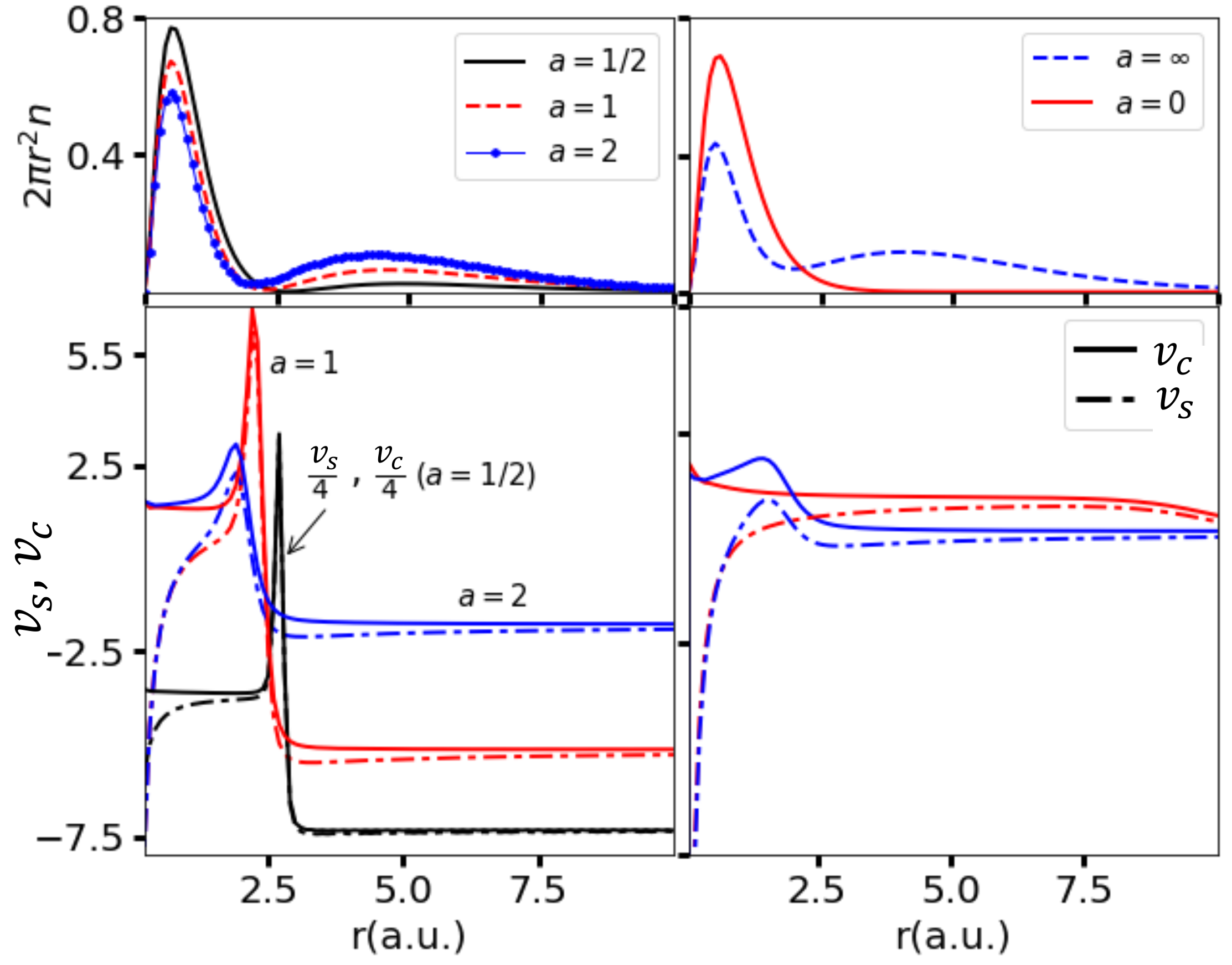}
\caption{Top panels: Left, initial densities for superposition states, 20:80 ($a=2$), 50:50 ($ a=1)$ and 80:20 ($a=1/2)$. Right, ground state ($a =0$) and excited state ($a=\infty$) densities. 
Lower panels: The corresponding KS and correlation potentials. }
\label{fig:vc_cmp}
\end{figure}

Taking different superpositions of the ground and excited states shows that the step and peak features are universally present in real 3D systems. Figure~\ref{fig:vc_cmp} shows the KS and correlation potentials at the initial time, when $a$ in Eq.~\ref{eq:intstate} is changed through $0,0.5,1, 2,\infty$ . We see that, for finite values of $a$,  as the fraction of excited state is increased the step and peak decrease in magnitude but extend over a larger region and move inward where more of the density is. The very sharp peak and large step seen when $a = 0.5$ (note it is scaled to fit on the plot) occur at a sharp minimum of the density and has a smaller impact on the ensuing dynamics than the softer but still prominent structures at large $a$ occurring in regions of greater density.  When the excited state is fully occupied ($a= \infty$) the KS potential is such to maintain the constant excited $^1P$ density at all times with a non-interacting doubly-occupied orbital, and the structure is not unlike that seen in the corresponding 1D excited helium atom of Ref.~\cite{EM12},  in both magnitude and shape; again, even the adiabatically-exact potential would have a completely different structure.

\section{True and Kohn-Sham Current}
\begin{figure}
\includegraphics[width=0.5\textwidth]{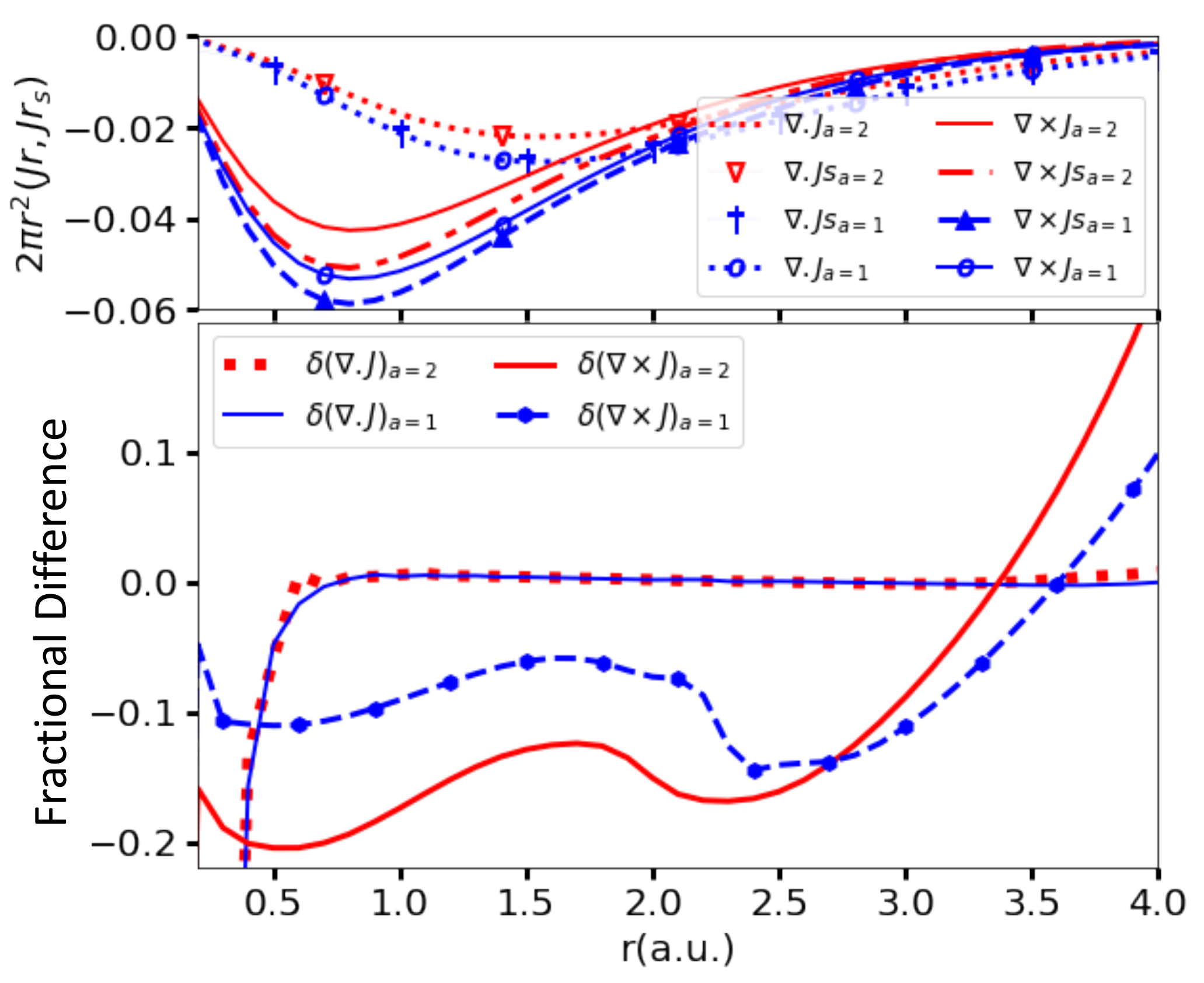}
\caption{Top panel: Divergence and the azimuthal component of the curl of the true and KS currents for $a=1$ and $a=2$. 
Lower panel: Fractional difference $\delta$ in the azimuthal component of curls and divergences of the current-densities along $\theta = 3\pi/4, t = T/8$ }
\label{fig:fracj}
\end{figure}

Finally, we ask, how closely does the exact KS system reproduce the exact interacting current in this case? It was recognized in the early days of TDDFT, that the exact KS current could differ from the true current by a rotational component~\cite{TK09b,MBAG02,GM12,AV05,SK16b}, but how large this difference could be for realistic systems in the non-perturbative regime was unknown. 
The KS and true currents are equal in their longitudinal component, thanks to the equation of continuity, $\nabla\cdot\bj = -\frac{\partial n}{\partial t}$, but they can differ in their rotational component. Indeed, for the two-electron singlet case with the KS system represented by a Slater determinant, it follows from Eq.~\ref{eqn:phi} that $\bj_s(\br,t) = n(\br,t)\nabla\alpha(\br,t)$. This implies that the true current-density would need to satisfy $\nabla \times (\bj/n) = 0$ in order for the KS current to possibly be equal to it. For our chosen $M=0$ superposition, the rotational component of the current (which has only an azimuthal component) is comparable to the longitudinal component (a movie is given in the Supplementary Material), and differs from the KS current. The fractional difference in the rotational component increases from about $10\%$  to  $20\%$ as we increase the proportion of the excited state in Eq.~(\ref{eq:intstate}) from $a = 1$ to $2$. This is shown in Fig.~\ref{fig:fracj}, where the fractional difference in the curl, $\delta(\nabla \times {\bf j}) = \frac{\nabla \times (\bj\s- \bj))}{\nabla\times\bj}$, is contrasted with that in the divergence $\delta(\nabla\cdot\bj) =\frac{\nabla\cdot(\bj\s -\bj)}{\partial n/\partial t}=\frac{\nabla\cdot(\bj\s +\partial n/\partial t)}{\partial n/\partial t}$; the latter comes only from numerical error, and is negligible except near the origin and at large $r$ where the denominator is very small. We note that the curl of the current is only non-zero in the azimuthal direction, and this is the component of the curl that is plotted in the figure. 

\section{Conclusion}
 
In summary, we have shown that the non-adiabatic dynamical features of the exact correlation potential, previously seen to arise in 1D model systems persist with comparable magnitudes in real 3D systems, and are not a consequence of reduced dimensionality. The results inform the on-going development of more accurate functionals in TDDFT  that capture these features~\cite{LM18,LM20b}, pressing the case to go beyond  adiabatic functionals. 
Hybrid functionals, including range-separated ones, where non-local density-dependence arises from the orbital dependence in exact exchange, do not capture these features;  this is particularly evident in the present two-electron case, where exact-exchange simply cancels the self-interaction in the Hartree potential. Furthermore, we have  demonstrated that the true interacting current differs from its KS counterpart, with the difference depending on the relative proportions of ground and excited state composing the state. The results thus advise caution when computing the current-density from the KS orbitals; this will be inherently approximate even if the exact xc functional was somehow known and used.
\acknowledgements{
Financial support from the National Science Foundation Award CHE-1940333 (DD)  and from the Department
of Energy, Office of Basic Energy Sciences, Division of Chemical
Sciences, Geosciences and Biosciences under Award No. DESC0020044 (NTM, LL) are gratefully acknowledged. J.F. acknowledges financial support from the European Research Council through grant ERC-2016-StG-714870, and by the Spanish Ministry for Science, Innovation, and Universities -- Agencia Estatal de Investigaci{\'o}n through grant RTI2018-099737-B-I00}

\bibliography{./ms.bib}

\begin{thebibliography}{61}%
\makeatletter
\providecommand \@ifxundefined [1]{%
 \@ifx{#1\undefined}
}%
\providecommand \@ifnum [1]{%
 \ifnum #1\expandafter \@firstoftwo
 \else \expandafter \@secondoftwo
 \fi
}%
\providecommand \@ifx [1]{%
 \ifx #1\expandafter \@firstoftwo
 \else \expandafter \@secondoftwo
 \fi
}%
\providecommand \natexlab [1]{#1}%
\providecommand \enquote  [1]{``#1''}%
\providecommand \bibnamefont  [1]{#1}%
\providecommand \bibfnamefont [1]{#1}%
\providecommand \citenamefont [1]{#1}%
\providecommand \href@noop [0]{\@secondoftwo}%
\providecommand \href [0]{\begingroup \@sanitize@url \@href}%
\providecommand \@href[1]{\@@startlink{#1}\@@href}%
\providecommand \@@href[1]{\endgroup#1\@@endlink}%
\providecommand \@sanitize@url [0]{\catcode `\\12\catcode `\$12\catcode
  `\&12\catcode `\#12\catcode `\^12\catcode `\_12\catcode `\%12\relax}%
\providecommand \@@startlink[1]{}%
\providecommand \@@endlink[0]{}%
\providecommand \url  [0]{\begingroup\@sanitize@url \@url }%
\providecommand \@url [1]{\endgroup\@href {#1}{\urlprefix }}%
\providecommand \urlprefix  [0]{URL }%
\providecommand \Eprint [0]{\href }%
\providecommand \doibase [0]{http://dx.doi.org/}%
\providecommand \selectlanguage [0]{\@gobble}%
\providecommand \bibinfo  [0]{\@secondoftwo}%
\providecommand \bibfield  [0]{\@secondoftwo}%
\providecommand \translation [1]{[#1]}%
\providecommand \BibitemOpen [0]{}%
\providecommand \bibitemStop [0]{}%
\providecommand \bibitemNoStop [0]{.\EOS\space}%
\providecommand \EOS [0]{\spacefactor3000\relax}%
\providecommand \BibitemShut  [1]{\csname bibitem#1\endcsname}%
\let\auto@bib@innerbib\@empty
\bibitem [{\citenamefont {Runge}\ and\ \citenamefont {Gross}(1984)}]{RG84}%
  \BibitemOpen
  \bibfield  {author} {\bibinfo {author} {\bibfnamefont {E.}~\bibnamefont
  {Runge}}\ and\ \bibinfo {author} {\bibfnamefont {E.~K.~U.}\ \bibnamefont
  {Gross}},\ }\href {\doibase 10.1103/PhysRevLett.52.997} {\bibfield  {journal}
  {\bibinfo  {journal} {Phys. Rev. Lett.}\ }\textbf {\bibinfo {volume} {52}},\
  \bibinfo {pages} {997} (\bibinfo {year} {1984})}\BibitemShut {NoStop}%
\bibitem [{\citenamefont {Maitra}(2016)}]{M16}%
  \BibitemOpen
  \bibfield  {author} {\bibinfo {author} {\bibfnamefont {N.~T.}\ \bibnamefont
  {Maitra}},\ }\href@noop {} {\bibfield  {journal} {\bibinfo  {journal} {J.
  Chem. Phys.}\ }\textbf {\bibinfo {volume} {144}},\ \bibinfo {pages} {220901}
  (\bibinfo {year} {2016})}\BibitemShut {NoStop}%
\bibitem [{\citenamefont {Marques}\ \emph {et~al.}(2012)\citenamefont
  {Marques}, \citenamefont {Maitra}, \citenamefont {Nogueira}, \citenamefont
  {Gross},\ and\ \citenamefont {Rubio}}]{TDDFTbook2012}%
  \BibitemOpen
  \bibinfo {editor} {\bibfnamefont {M.~A.}\ \bibnamefont {Marques}}, \bibinfo
  {editor} {\bibfnamefont {N.~T.}\ \bibnamefont {Maitra}}, \bibinfo {editor}
  {\bibfnamefont {F.~M.}\ \bibnamefont {Nogueira}}, \bibinfo {editor}
  {\bibfnamefont {E.~K.}\ \bibnamefont {Gross}}, \ and\ \bibinfo {editor}
  {\bibfnamefont {A.}~\bibnamefont {Rubio}},\ eds.,\ \href@noop {} {\emph
  {\bibinfo {title} {Fundamentals of time-dependent density functional
  theory}}},\ Vol.\ \bibinfo {volume} {837}\ (\bibinfo  {publisher}
  {Springer},\ \bibinfo {year} {2012})\BibitemShut {NoStop}%
\bibitem [{\citenamefont {Ullrich}(2011)}]{Carstenbook}%
  \BibitemOpen
  \bibfield  {author} {\bibinfo {author} {\bibfnamefont {C.~A.}\ \bibnamefont
  {Ullrich}},\ }\href@noop {} {\emph {\bibinfo {title} {Time-dependent
  density-functional theory: concepts and applications}}}\ (\bibinfo
  {publisher} {Oxford University Press},\ \bibinfo {year} {2011})\BibitemShut
  {NoStop}%
\bibitem [{\citenamefont {Draeger}\ \emph {et~al.}(2017)\citenamefont
  {Draeger}, \citenamefont {Andrade}, \citenamefont {Gunnels}, \citenamefont
  {Bhatele}, \citenamefont {Schleife},\ and\ \citenamefont
  {Correa}}]{DAGBSC17}%
  \BibitemOpen
  \bibfield  {author} {\bibinfo {author} {\bibfnamefont {E.~W.}\ \bibnamefont
  {Draeger}}, \bibinfo {author} {\bibfnamefont {X.}~\bibnamefont {Andrade}},
  \bibinfo {author} {\bibfnamefont {J.~A.}\ \bibnamefont {Gunnels}}, \bibinfo
  {author} {\bibfnamefont {A.}~\bibnamefont {Bhatele}}, \bibinfo {author}
  {\bibfnamefont {A.}~\bibnamefont {Schleife}}, \ and\ \bibinfo {author}
  {\bibfnamefont {A.~A.}\ \bibnamefont {Correa}},\ }\href {\doibase
  https://doi.org/10.1016/j.jpdc.2017.02.005} {\bibfield  {journal} {\bibinfo
  {journal} {Journal of Parallel and Distributed Computing}\ }\textbf {\bibinfo
  {volume} {106}},\ \bibinfo {pages} {205 } (\bibinfo {year}
  {2017})}\BibitemShut {NoStop}%
\bibitem [{\citenamefont {Sato}(2021)}]{S21}%
  \BibitemOpen
  \bibfield  {author} {\bibinfo {author} {\bibfnamefont {S.~A.}\ \bibnamefont
  {Sato}},\ }\href {\doibase https://doi.org/10.1016/j.commatsci.2020.110274}
  {\bibfield  {journal} {\bibinfo  {journal} {Computational Materials Science}\
  ,\ \bibinfo {pages} {110274}} (\bibinfo {year} {2021})}\BibitemShut {NoStop}%
\bibitem [{\citenamefont {Guandalini}\ \emph {et~al.}(2021)\citenamefont
  {Guandalini}, \citenamefont {Cocchi}, \citenamefont {Pittalis}, \citenamefont
  {Ruini},\ and\ \citenamefont {Rozzi}}]{GCPRR20}%
  \BibitemOpen
  \bibfield  {author} {\bibinfo {author} {\bibfnamefont {A.}~\bibnamefont
  {Guandalini}}, \bibinfo {author} {\bibfnamefont {C.}~\bibnamefont {Cocchi}},
  \bibinfo {author} {\bibfnamefont {S.}~\bibnamefont {Pittalis}}, \bibinfo
  {author} {\bibfnamefont {A.}~\bibnamefont {Ruini}}, \ and\ \bibinfo {author}
  {\bibfnamefont {C.~A.}\ \bibnamefont {Rozzi}},\ }\href {\doibase
  10.1039/D0CP04958A} {\bibfield  {journal} {\bibinfo  {journal} {Phys. Chem.
  Chem. Phys.}\ ,\ } (\bibinfo {year} {2021})}\BibitemShut {NoStop}%
\bibitem [{\citenamefont {Singh}\ \emph {et~al.}(2021)\citenamefont {Singh},
  \citenamefont {Elliott}, \citenamefont {Dewhurst},\ and\ \citenamefont
  {Sharma}}]{SEDS21}%
  \BibitemOpen
  \bibfield  {author} {\bibinfo {author} {\bibfnamefont {N.}~\bibnamefont
  {Singh}}, \bibinfo {author} {\bibfnamefont {P.}~\bibnamefont {Elliott}},
  \bibinfo {author} {\bibfnamefont {J.~K.}\ \bibnamefont {Dewhurst}}, \ and\
  \bibinfo {author} {\bibfnamefont {S.}~\bibnamefont {Sharma}},\ }\href
  {\doibase 10.1103/PhysRevB.103.134402} {\bibfield  {journal} {\bibinfo
  {journal} {Phys. Rev. B}\ }\textbf {\bibinfo {volume} {103}},\ \bibinfo
  {pages} {134402} (\bibinfo {year} {2021})}\BibitemShut {NoStop}%
\bibitem [{\citenamefont {Folorunso}\ \emph {et~al.}(2021)\citenamefont
  {Folorunso}, \citenamefont {Bruner}, \citenamefont {Mauger}, \citenamefont
  {Hamer}, \citenamefont {Hernandez}, \citenamefont {Jones}, \citenamefont
  {DiMauro}, \citenamefont {Gaarde}, \citenamefont {Schafer},\ and\
  \citenamefont {Lopata}}]{FBMHH21}%
  \BibitemOpen
  \bibfield  {author} {\bibinfo {author} {\bibfnamefont {A.~S.}\ \bibnamefont
  {Folorunso}}, \bibinfo {author} {\bibfnamefont {A.}~\bibnamefont {Bruner}},
  \bibinfo {author} {\bibfnamefont {F.~m.~c.}\ \bibnamefont {Mauger}}, \bibinfo
  {author} {\bibfnamefont {K.~A.}\ \bibnamefont {Hamer}}, \bibinfo {author}
  {\bibfnamefont {S.}~\bibnamefont {Hernandez}}, \bibinfo {author}
  {\bibfnamefont {R.~R.}\ \bibnamefont {Jones}}, \bibinfo {author}
  {\bibfnamefont {L.~F.}\ \bibnamefont {DiMauro}}, \bibinfo {author}
  {\bibfnamefont {M.~B.}\ \bibnamefont {Gaarde}}, \bibinfo {author}
  {\bibfnamefont {K.~J.}\ \bibnamefont {Schafer}}, \ and\ \bibinfo {author}
  {\bibfnamefont {K.}~\bibnamefont {Lopata}},\ }\href {\doibase
  10.1103/PhysRevLett.126.133002} {\bibfield  {journal} {\bibinfo  {journal}
  {Phys. Rev. Lett.}\ }\textbf {\bibinfo {volume} {126}},\ \bibinfo {pages}
  {133002} (\bibinfo {year} {2021})}\BibitemShut {NoStop}%
\bibitem [{\citenamefont {Yost}\ \emph {et~al.}(2017)\citenamefont {Yost},
  \citenamefont {Yao},\ and\ \citenamefont {Kanai}}]{YYK17}%
  \BibitemOpen
  \bibfield  {author} {\bibinfo {author} {\bibfnamefont {D.~C.}\ \bibnamefont
  {Yost}}, \bibinfo {author} {\bibfnamefont {Y.}~\bibnamefont {Yao}}, \ and\
  \bibinfo {author} {\bibfnamefont {Y.}~\bibnamefont {Kanai}},\ }\href
  {\doibase 10.1103/PhysRevB.96.115134} {\bibfield  {journal} {\bibinfo
  {journal} {Phys. Rev. B}\ }\textbf {\bibinfo {volume} {96}},\ \bibinfo
  {pages} {115134} (\bibinfo {year} {2017})}\BibitemShut {NoStop}%
\bibitem [{\citenamefont {Raghunathan}\ and\ \citenamefont
  {Nest}(2011)}]{RN11}%
  \BibitemOpen
  \bibfield  {author} {\bibinfo {author} {\bibfnamefont {S.}~\bibnamefont
  {Raghunathan}}\ and\ \bibinfo {author} {\bibfnamefont {M.}~\bibnamefont
  {Nest}},\ }\href {\doibase 10.1021/ct200270t} {\bibfield  {journal} {\bibinfo
   {journal} {J. Chem. Theory and Comput.}\ }\textbf {\bibinfo {volume} {7}},\
  \bibinfo {pages} {2492} (\bibinfo {year} {2011})}\BibitemShut {NoStop}%
\bibitem [{\citenamefont {Ramakrishnan}\ and\ \citenamefont
  {Nest}(2012)}]{RN12}%
  \BibitemOpen
  \bibfield  {author} {\bibinfo {author} {\bibfnamefont {R.}~\bibnamefont
  {Ramakrishnan}}\ and\ \bibinfo {author} {\bibfnamefont {M.}~\bibnamefont
  {Nest}},\ }\href {\doibase 10.1103/PhysRevA.85.054501} {\bibfield  {journal}
  {\bibinfo  {journal} {Phys. Rev. A}\ }\textbf {\bibinfo {volume} {85}},\
  \bibinfo {pages} {054501} (\bibinfo {year} {2012})}\BibitemShut {NoStop}%
\bibitem [{\citenamefont {Habenicht}\ \emph {et~al.}(2014)\citenamefont
  {Habenicht}, \citenamefont {Tani}, \citenamefont {Provorse},\ and\
  \citenamefont {Isborn}}]{HTPI14}%
  \BibitemOpen
  \bibfield  {author} {\bibinfo {author} {\bibfnamefont {B.~F.}\ \bibnamefont
  {Habenicht}}, \bibinfo {author} {\bibfnamefont {N.~P.}\ \bibnamefont {Tani}},
  \bibinfo {author} {\bibfnamefont {M.~R.}\ \bibnamefont {Provorse}}, \ and\
  \bibinfo {author} {\bibfnamefont {C.~M.}\ \bibnamefont {Isborn}},\
  }\href@noop {} {\bibfield  {journal} {\bibinfo  {journal} {J. Chem. Phys.}\
  }\textbf {\bibinfo {volume} {141}},\ \bibinfo {pages} {184112} (\bibinfo
  {year} {2014})}\BibitemShut {NoStop}%
\bibitem [{\citenamefont {Wijewardane}\ and\ \citenamefont
  {Ullrich}(2008)}]{WU08}%
  \BibitemOpen
  \bibfield  {author} {\bibinfo {author} {\bibfnamefont {H.~O.}\ \bibnamefont
  {Wijewardane}}\ and\ \bibinfo {author} {\bibfnamefont {C.~A.}\ \bibnamefont
  {Ullrich}},\ }\href {\doibase 10.1103/PhysRevLett.100.056404} {\bibfield
  {journal} {\bibinfo  {journal} {Phys. Rev. Lett.}\ }\textbf {\bibinfo
  {volume} {100}},\ \bibinfo {pages} {056404} (\bibinfo {year}
  {2008})}\BibitemShut {NoStop}%
\bibitem [{\citenamefont {Gao}\ \emph {et~al.}(2017)\citenamefont {Gao},
  \citenamefont {Dinh}, \citenamefont {Reinhard},\ and\ \citenamefont
  {Suraud}}]{GDRS17}%
  \BibitemOpen
  \bibfield  {author} {\bibinfo {author} {\bibfnamefont {C.-Z.}\ \bibnamefont
  {Gao}}, \bibinfo {author} {\bibfnamefont {P.~M.}\ \bibnamefont {Dinh}},
  \bibinfo {author} {\bibfnamefont {P.-G.}\ \bibnamefont {Reinhard}}, \ and\
  \bibinfo {author} {\bibfnamefont {E.}~\bibnamefont {Suraud}},\ }\href@noop {}
  {\bibfield  {journal} {\bibinfo  {journal} {Phys. Chem. Chem. Phys.}\
  }\textbf {\bibinfo {volume} {19}},\ \bibinfo {pages} {19784} (\bibinfo {year}
  {2017})}\BibitemShut {NoStop}%
\bibitem [{\citenamefont {Bostr\"om}\ \emph {et~al.}(2018)\citenamefont
  {Bostr\"om}, \citenamefont {Mikkelsen}, \citenamefont {Verdozzi},
  \citenamefont {Perfetto},\ and\ \citenamefont {Stefanucci}}]{BMVPS18}%
  \BibitemOpen
  \bibfield  {author} {\bibinfo {author} {\bibfnamefont {E.~V.}\ \bibnamefont
  {Bostr\"om}}, \bibinfo {author} {\bibfnamefont {A.}~\bibnamefont
  {Mikkelsen}}, \bibinfo {author} {\bibfnamefont {C.}~\bibnamefont {Verdozzi}},
  \bibinfo {author} {\bibfnamefont {E.}~\bibnamefont {Perfetto}}, \ and\
  \bibinfo {author} {\bibfnamefont {G.}~\bibnamefont {Stefanucci}},\ }\href
  {\doibase 10.1021/acs.nanolett.7b03995} {\bibfield  {journal} {\bibinfo
  {journal} {Nano Letters}\ }\textbf {\bibinfo {volume} {18}},\ \bibinfo
  {pages} {785} (\bibinfo {year} {2018})}\BibitemShut {NoStop}%
\bibitem [{\citenamefont {Krumland}\ \emph {et~al.}(2020)\citenamefont
  {Krumland}, \citenamefont {Valencia}, \citenamefont {Pittalis}, \citenamefont
  {Rozzi},\ and\ \citenamefont {Cocchi}}]{KVPRC20}%
  \BibitemOpen
  \bibfield  {author} {\bibinfo {author} {\bibfnamefont {J.}~\bibnamefont
  {Krumland}}, \bibinfo {author} {\bibfnamefont {A.~M.}\ \bibnamefont
  {Valencia}}, \bibinfo {author} {\bibfnamefont {S.}~\bibnamefont {Pittalis}},
  \bibinfo {author} {\bibfnamefont {C.~A.}\ \bibnamefont {Rozzi}}, \ and\
  \bibinfo {author} {\bibfnamefont {C.}~\bibnamefont {Cocchi}},\ }\href
  {\doibase 10.1063/5.0008194} {\bibfield  {journal} {\bibinfo  {journal} {The
  Journal of Chemical Physics}\ }\textbf {\bibinfo {volume} {153}},\ \bibinfo
  {pages} {054106} (\bibinfo {year} {2020})},\ \Eprint
  {http://arxiv.org/abs/https://doi.org/10.1063/5.0008194}
  {https://doi.org/10.1063/5.0008194} \BibitemShut {NoStop}%
\bibitem [{\citenamefont {Quashie}\ \emph {et~al.}(2017)\citenamefont
  {Quashie}, \citenamefont {Saha}, \citenamefont {Andrade},\ and\ \citenamefont
  {Correa}}]{QSAC17}%
  \BibitemOpen
  \bibfield  {author} {\bibinfo {author} {\bibfnamefont {E.~E.}\ \bibnamefont
  {Quashie}}, \bibinfo {author} {\bibfnamefont {B.~C.}\ \bibnamefont {Saha}},
  \bibinfo {author} {\bibfnamefont {X.}~\bibnamefont {Andrade}}, \ and\
  \bibinfo {author} {\bibfnamefont {A.~A.}\ \bibnamefont {Correa}},\ }\href
  {\doibase 10.1103/PhysRevA.95.042517} {\bibfield  {journal} {\bibinfo
  {journal} {Phys. Rev. A}\ }\textbf {\bibinfo {volume} {95}},\ \bibinfo
  {pages} {042517} (\bibinfo {year} {2017})}\BibitemShut {NoStop}%
\bibitem [{\citenamefont {Gao}\ \emph {et~al.}(2014)\citenamefont {Gao},
  \citenamefont {Wang}, \citenamefont {Wang},\ and\ \citenamefont
  {Zhang}}]{GWWZ14}%
  \BibitemOpen
  \bibfield  {author} {\bibinfo {author} {\bibfnamefont {C.-Z.}\ \bibnamefont
  {Gao}}, \bibinfo {author} {\bibfnamefont {J.}~\bibnamefont {Wang}}, \bibinfo
  {author} {\bibfnamefont {F.}~\bibnamefont {Wang}}, \ and\ \bibinfo {author}
  {\bibfnamefont {F.-S.}\ \bibnamefont {Zhang}},\ }\href {\doibase
  10.1063/1.4863635} {\bibfield  {journal} {\bibinfo  {journal} {The Journal of
  Chemical Physics}\ }\textbf {\bibinfo {volume} {140}},\ \bibinfo {pages}
  {054308} (\bibinfo {year} {2014})}\BibitemShut {NoStop}%
\bibitem [{\citenamefont {Henkel}\ \emph {et~al.}(2009)\citenamefont {Henkel},
  \citenamefont {Keim}, \citenamefont {L\"udde},\ and\ \citenamefont
  {Kirchner}}]{HKLK09}%
  \BibitemOpen
  \bibfield  {author} {\bibinfo {author} {\bibfnamefont {N.}~\bibnamefont
  {Henkel}}, \bibinfo {author} {\bibfnamefont {M.}~\bibnamefont {Keim}},
  \bibinfo {author} {\bibfnamefont {H.~J.}\ \bibnamefont {L\"udde}}, \ and\
  \bibinfo {author} {\bibfnamefont {T.}~\bibnamefont {Kirchner}},\ }\href
  {\doibase 10.1103/PhysRevA.80.032704} {\bibfield  {journal} {\bibinfo
  {journal} {Phys. Rev. A}\ }\textbf {\bibinfo {volume} {80}},\ \bibinfo
  {pages} {032704} (\bibinfo {year} {2009})}\BibitemShut {NoStop}%
\bibitem [{\citenamefont {Da}\ \emph {et~al.}(2017)\citenamefont {Da},
  \citenamefont {Liu}, \citenamefont {Yamamoto}, \citenamefont {Ueda},
  \citenamefont {Watanabe}, \citenamefont {Cuong}, \citenamefont {Li},
  \citenamefont {Tsukagoshi}, \citenamefont {Yoshikawa}, \citenamefont {Iwai},
  \citenamefont {Tanuma}, \citenamefont {Guo}, \citenamefont {Gao},
  \citenamefont {Sun},\ and\ \citenamefont {Ding}}]{DLYU17}%
  \BibitemOpen
  \bibfield  {author} {\bibinfo {author} {\bibfnamefont {B.}~\bibnamefont
  {Da}}, \bibinfo {author} {\bibfnamefont {J.}~\bibnamefont {Liu}}, \bibinfo
  {author} {\bibfnamefont {M.}~\bibnamefont {Yamamoto}}, \bibinfo {author}
  {\bibfnamefont {Y.}~\bibnamefont {Ueda}}, \bibinfo {author} {\bibfnamefont
  {K.}~\bibnamefont {Watanabe}}, \bibinfo {author} {\bibfnamefont {N.~T.}\
  \bibnamefont {Cuong}}, \bibinfo {author} {\bibfnamefont {S.}~\bibnamefont
  {Li}}, \bibinfo {author} {\bibfnamefont {K.}~\bibnamefont {Tsukagoshi}},
  \bibinfo {author} {\bibfnamefont {H.}~\bibnamefont {Yoshikawa}}, \bibinfo
  {author} {\bibfnamefont {H.}~\bibnamefont {Iwai}}, \bibinfo {author}
  {\bibfnamefont {S.}~\bibnamefont {Tanuma}}, \bibinfo {author} {\bibfnamefont
  {H.}~\bibnamefont {Guo}}, \bibinfo {author} {\bibfnamefont {Z.}~\bibnamefont
  {Gao}}, \bibinfo {author} {\bibfnamefont {X.}~\bibnamefont {Sun}}, \ and\
  \bibinfo {author} {\bibfnamefont {Z.}~\bibnamefont {Ding}},\ }\href@noop {}
  {\bibfield  {journal} {\bibinfo  {journal} {Nat. Commun.}\ }\textbf {\bibinfo
  {volume} {8}},\ \bibinfo {pages} {15629} (\bibinfo {year}
  {2017})}\BibitemShut {NoStop}%
\bibitem [{\citenamefont {Yao}\ \emph {et~al.}(2021)\citenamefont {Yao},
  \citenamefont {Lee}, \citenamefont {Ceresoli},\ and\ \citenamefont
  {Cho}}]{YYCC21}%
  \BibitemOpen
  \bibfield  {author} {\bibinfo {author} {\bibfnamefont {X.}~\bibnamefont
  {Yao}}, \bibinfo {author} {\bibfnamefont {Y.}~\bibnamefont {Lee}}, \bibinfo
  {author} {\bibfnamefont {D.}~\bibnamefont {Ceresoli}}, \ and\ \bibinfo
  {author} {\bibfnamefont {K.}~\bibnamefont {Cho}},\ }\href {\doibase
  10.1021/acs.jpca.0c11309} {\bibfield  {journal} {\bibinfo  {journal} {The
  Journal of Physical Chemistry A}\ }\textbf {\bibinfo {volume} {125}},\
  \bibinfo {pages} {4524} (\bibinfo {year} {2021})},\ \bibinfo {note} {pMID:
  34019398},\ \Eprint
  {http://arxiv.org/abs/https://doi.org/10.1021/acs.jpca.0c11309}
  {https://doi.org/10.1021/acs.jpca.0c11309} \BibitemShut {NoStop}%
\bibitem [{\citenamefont {Lacombe}\ and\ \citenamefont {Maitra}(2020)}]{LM20b}%
  \BibitemOpen
  \bibfield  {author} {\bibinfo {author} {\bibfnamefont {L.}~\bibnamefont
  {Lacombe}}\ and\ \bibinfo {author} {\bibfnamefont {N.~T.}\ \bibnamefont
  {Maitra}},\ }\href {\doibase 10.1039/D0FD00049C} {\bibfield  {journal}
  {\bibinfo  {journal} {Faraday Discuss.}\ }\textbf {\bibinfo {volume} {224}},\
  \bibinfo {pages} {382} (\bibinfo {year} {2020})}\BibitemShut {NoStop}%
\bibitem [{\citenamefont {Luo}\ \emph {et~al.}(2014)\citenamefont {Luo},
  \citenamefont {Fuks}, \citenamefont {Sandoval}, \citenamefont {Elliott},\
  and\ \citenamefont {Maitra}}]{LFSEM14}%
  \BibitemOpen
  \bibfield  {author} {\bibinfo {author} {\bibfnamefont {K.}~\bibnamefont
  {Luo}}, \bibinfo {author} {\bibfnamefont {J.~I.}\ \bibnamefont {Fuks}},
  \bibinfo {author} {\bibfnamefont {E.~D.}\ \bibnamefont {Sandoval}}, \bibinfo
  {author} {\bibfnamefont {P.}~\bibnamefont {Elliott}}, \ and\ \bibinfo
  {author} {\bibfnamefont {N.~T.}\ \bibnamefont {Maitra}},\ }\href@noop {}
  {\bibfield  {journal} {\bibinfo  {journal} {J. Chem. Phys}\ }\textbf
  {\bibinfo {volume} {140}},\ \bibinfo {pages} {18A515} (\bibinfo {year}
  {2014})}\BibitemShut {NoStop}%
\bibitem [{\citenamefont {Suzuki}\ \emph {et~al.}(2017)\citenamefont {Suzuki},
  \citenamefont {Lacombe}, \citenamefont {Watanabe},\ and\ \citenamefont
  {Maitra}}]{SLWM17}%
  \BibitemOpen
  \bibfield  {author} {\bibinfo {author} {\bibfnamefont {Y.}~\bibnamefont
  {Suzuki}}, \bibinfo {author} {\bibfnamefont {L.}~\bibnamefont {Lacombe}},
  \bibinfo {author} {\bibfnamefont {K.}~\bibnamefont {Watanabe}}, \ and\
  \bibinfo {author} {\bibfnamefont {N.~T.}\ \bibnamefont {Maitra}},\ }\href
  {\doibase 10.1103/PhysRevLett.119.263401} {\bibfield  {journal} {\bibinfo
  {journal} {Phys. Rev. Lett.}\ }\textbf {\bibinfo {volume} {119}},\ \bibinfo
  {pages} {263401} (\bibinfo {year} {2017})}\BibitemShut {NoStop}%
\bibitem [{\citenamefont {Elliott}\ \emph {et~al.}(2012)\citenamefont
  {Elliott}, \citenamefont {Fuks}, \citenamefont {Rubio},\ and\ \citenamefont
  {Maitra}}]{EFRM12}%
  \BibitemOpen
  \bibfield  {author} {\bibinfo {author} {\bibfnamefont {P.}~\bibnamefont
  {Elliott}}, \bibinfo {author} {\bibfnamefont {J.~I.}\ \bibnamefont {Fuks}},
  \bibinfo {author} {\bibfnamefont {A.}~\bibnamefont {Rubio}}, \ and\ \bibinfo
  {author} {\bibfnamefont {N.~T.}\ \bibnamefont {Maitra}},\ }\href {\doibase
  10.1103/PhysRevLett.109.266404} {\bibfield  {journal} {\bibinfo  {journal}
  {Phys. Rev. Lett.}\ }\textbf {\bibinfo {volume} {109}},\ \bibinfo {pages}
  {266404} (\bibinfo {year} {2012})}\BibitemShut {NoStop}%
\bibitem [{\citenamefont {Fuks}\ \emph {et~al.}(2013)\citenamefont {Fuks},
  \citenamefont {Elliott}, \citenamefont {Rubio},\ and\ \citenamefont
  {Maitra}}]{FERM13}%
  \BibitemOpen
  \bibfield  {author} {\bibinfo {author} {\bibfnamefont {J.~I.}\ \bibnamefont
  {Fuks}}, \bibinfo {author} {\bibfnamefont {P.}~\bibnamefont {Elliott}},
  \bibinfo {author} {\bibfnamefont {A.}~\bibnamefont {Rubio}}, \ and\ \bibinfo
  {author} {\bibfnamefont {N.~T.}\ \bibnamefont {Maitra}},\ }\href@noop {}
  {\bibfield  {journal} {\bibinfo  {journal} {J. Phys. Chem. Lett.}\ }\textbf
  {\bibinfo {volume} {4}},\ \bibinfo {pages} {735} (\bibinfo {year}
  {2013})}\BibitemShut {NoStop}%
\bibitem [{\citenamefont {Lacombe}\ and\ \citenamefont {Maitra}(2019)}]{LM18}%
  \BibitemOpen
  \bibfield  {author} {\bibinfo {author} {\bibfnamefont {L.}~\bibnamefont
  {Lacombe}}\ and\ \bibinfo {author} {\bibfnamefont {N.~T.}\ \bibnamefont
  {Maitra}},\ }\href {\doibase 10.1021/acs.jctc.8b01159} {\bibfield  {journal}
  {\bibinfo  {journal} {Journal of Chemical Theory and Computation}\ }\textbf
  {\bibinfo {volume} {15}},\ \bibinfo {pages} {1672} (\bibinfo {year}
  {2019})},\ \Eprint
  {http://arxiv.org/abs/https://doi.org/10.1021/acs.jctc.8b01159}
  {https://doi.org/10.1021/acs.jctc.8b01159} \BibitemShut {NoStop}%
\bibitem [{\citenamefont {Lacombe}\ \emph {et~al.}(2018)\citenamefont
  {Lacombe}, \citenamefont {Suzuki}, \citenamefont {Watanabe},\ and\
  \citenamefont {Maitra}}]{LSWM18}%
  \BibitemOpen
  \bibfield  {author} {\bibinfo {author} {\bibfnamefont {L.}~\bibnamefont
  {Lacombe}}, \bibinfo {author} {\bibfnamefont {Y.}~\bibnamefont {Suzuki}},
  \bibinfo {author} {\bibfnamefont {K.}~\bibnamefont {Watanabe}}, \ and\
  \bibinfo {author} {\bibfnamefont {N.~T.}\ \bibnamefont {Maitra}},\ }\href
  {\doibase 10.1140/epjb/e2018-90101-2} {\bibfield  {journal} {\bibinfo
  {journal} {The European Physical Journal B}\ }\textbf {\bibinfo {volume}
  {91}},\ \bibinfo {pages} {96} (\bibinfo {year} {2018})}\BibitemShut {NoStop}%
\bibitem [{\citenamefont {Ramsden}\ and\ \citenamefont {Godby}(2012)}]{RG12}%
  \BibitemOpen
  \bibfield  {author} {\bibinfo {author} {\bibfnamefont {J.}~\bibnamefont
  {Ramsden}}\ and\ \bibinfo {author} {\bibfnamefont {R.}~\bibnamefont
  {Godby}},\ }\href@noop {} {\bibfield  {journal} {\bibinfo  {journal} {Phys.
  Rev. Lett.}\ }\textbf {\bibinfo {volume} {109}},\ \bibinfo {pages} {036402}
  (\bibinfo {year} {2012})}\BibitemShut {NoStop}%
\bibitem [{\citenamefont {Fuks}\ \emph {et~al.}(2016)\citenamefont {Fuks},
  \citenamefont {Nielsen}, \citenamefont {Ruggenthaler},\ and\ \citenamefont
  {Maitra}}]{FNRM16}%
  \BibitemOpen
  \bibfield  {author} {\bibinfo {author} {\bibfnamefont {J.~I.}\ \bibnamefont
  {Fuks}}, \bibinfo {author} {\bibfnamefont {S.}~\bibnamefont {Nielsen}},
  \bibinfo {author} {\bibfnamefont {M.}~\bibnamefont {Ruggenthaler}}, \ and\
  \bibinfo {author} {\bibfnamefont {N.~T.}\ \bibnamefont {Maitra}},\
  }\href@noop {} {\bibfield  {journal} {\bibinfo  {journal} {Phys. Chem. Chem.
  Phys.}\ }\textbf {\bibinfo {volume} {18}},\ \bibinfo {pages} {20976}
  (\bibinfo {year} {2016})}\BibitemShut {NoStop}%
\bibitem [{\citenamefont {Fuks}\ \emph {et~al.}(2018)\citenamefont {Fuks},
  \citenamefont {Lacombe}, \citenamefont {Nielsen},\ and\ \citenamefont
  {Maitra}}]{FLNM18}%
  \BibitemOpen
  \bibfield  {author} {\bibinfo {author} {\bibfnamefont {J.~I.}\ \bibnamefont
  {Fuks}}, \bibinfo {author} {\bibfnamefont {L.}~\bibnamefont {Lacombe}},
  \bibinfo {author} {\bibfnamefont {S.~E.~B.}\ \bibnamefont {Nielsen}}, \ and\
  \bibinfo {author} {\bibfnamefont {N.~T.}\ \bibnamefont {Maitra}},\ }\href
  {\doibase 10.1039/C8CP03957G} {\bibfield  {journal} {\bibinfo  {journal}
  {Phys. Chem. Chem. Phys.}\ }\textbf {\bibinfo {volume} {20}},\ \bibinfo
  {pages} {26145} (\bibinfo {year} {2018})}\BibitemShut {NoStop}%
\bibitem [{\citenamefont {Hodgson}\ \emph {et~al.}(2016)\citenamefont
  {Hodgson}, \citenamefont {Ramsden},\ and\ \citenamefont {Godby}}]{HRG16}%
  \BibitemOpen
  \bibfield  {author} {\bibinfo {author} {\bibfnamefont {M.~J.~P.}\
  \bibnamefont {Hodgson}}, \bibinfo {author} {\bibfnamefont {J.~D.}\
  \bibnamefont {Ramsden}}, \ and\ \bibinfo {author} {\bibfnamefont {R.~W.}\
  \bibnamefont {Godby}},\ }\href {\doibase 10.1103/PhysRevB.93.155146}
  {\bibfield  {journal} {\bibinfo  {journal} {Phys. Rev. B}\ }\textbf {\bibinfo
  {volume} {93}},\ \bibinfo {pages} {155146} (\bibinfo {year}
  {2016})}\BibitemShut {NoStop}%
\bibitem [{\citenamefont {Hodgson}\ \emph {et~al.}(2013)\citenamefont
  {Hodgson}, \citenamefont {Ramsden}, \citenamefont {Chapman}, \citenamefont
  {Lillystone},\ and\ \citenamefont {Godby}}]{HRCLG13}%
  \BibitemOpen
  \bibfield  {author} {\bibinfo {author} {\bibfnamefont {M.~J.~P.}\
  \bibnamefont {Hodgson}}, \bibinfo {author} {\bibfnamefont {J.~D.}\
  \bibnamefont {Ramsden}}, \bibinfo {author} {\bibfnamefont {J.~B.~J.}\
  \bibnamefont {Chapman}}, \bibinfo {author} {\bibfnamefont {P.}~\bibnamefont
  {Lillystone}}, \ and\ \bibinfo {author} {\bibfnamefont {R.~W.}\ \bibnamefont
  {Godby}},\ }\href {\doibase 10.1103/PhysRevB.88.241102} {\bibfield  {journal}
  {\bibinfo  {journal} {Phys. Rev. B}\ }\textbf {\bibinfo {volume} {88}},\
  \bibinfo {pages} {241102} (\bibinfo {year} {2013})}\BibitemShut {NoStop}%
\bibitem [{\citenamefont {D'Agosta}\ and\ \citenamefont
  {Vignale}(2005)}]{AV05}%
  \BibitemOpen
  \bibfield  {author} {\bibinfo {author} {\bibfnamefont {R.}~\bibnamefont
  {D'Agosta}}\ and\ \bibinfo {author} {\bibfnamefont {G.}~\bibnamefont
  {Vignale}},\ }\href {\doibase 10.1103/PhysRevB.71.245103} {\bibfield
  {journal} {\bibinfo  {journal} {Phys. Rev. B}\ }\textbf {\bibinfo {volume}
  {71}},\ \bibinfo {pages} {245103} (\bibinfo {year} {2005})}\BibitemShut
  {NoStop}%
\bibitem [{\citenamefont {Ruggenthaler}\ \emph {et~al.}(2013)\citenamefont
  {Ruggenthaler}, \citenamefont {Nielsen},\ and\ \citenamefont {van
  Leeuwen}}]{RNL13}%
  \BibitemOpen
  \bibfield  {author} {\bibinfo {author} {\bibfnamefont {M.}~\bibnamefont
  {Ruggenthaler}}, \bibinfo {author} {\bibfnamefont {S.~E.~B.}\ \bibnamefont
  {Nielsen}}, \ and\ \bibinfo {author} {\bibfnamefont {R.}~\bibnamefont {van
  Leeuwen}},\ }\href {\doibase 10.1103/PhysRevA.88.022512} {\bibfield
  {journal} {\bibinfo  {journal} {Phys. Rev. A}\ }\textbf {\bibinfo {volume}
  {88}},\ \bibinfo {pages} {022512} (\bibinfo {year} {2013})}\BibitemShut
  {NoStop}%
\bibitem [{\citenamefont {Elliott}\ and\ \citenamefont {Maitra}(2012)}]{EM12}%
  \BibitemOpen
  \bibfield  {author} {\bibinfo {author} {\bibfnamefont {P.}~\bibnamefont
  {Elliott}}\ and\ \bibinfo {author} {\bibfnamefont {N.~T.}\ \bibnamefont
  {Maitra}},\ }\href {\doibase 10.1103/PhysRevA.85.052510} {\bibfield
  {journal} {\bibinfo  {journal} {Phys. Rev. A}\ }\textbf {\bibinfo {volume}
  {85}},\ \bibinfo {pages} {052510} (\bibinfo {year} {2012})}\BibitemShut
  {NoStop}%
\bibitem [{\citenamefont {Thiele}\ \emph {et~al.}(2008)\citenamefont {Thiele},
  \citenamefont {Gross},\ and\ \citenamefont {K\"ummel}}]{TGK08}%
  \BibitemOpen
  \bibfield  {author} {\bibinfo {author} {\bibfnamefont {M.}~\bibnamefont
  {Thiele}}, \bibinfo {author} {\bibfnamefont {E.~K.~U.}\ \bibnamefont
  {Gross}}, \ and\ \bibinfo {author} {\bibfnamefont {S.}~\bibnamefont
  {K\"ummel}},\ }\href {\doibase 10.1103/PhysRevLett.100.153004} {\bibfield
  {journal} {\bibinfo  {journal} {Phys. Rev. Lett.}\ }\textbf {\bibinfo
  {volume} {100}},\ \bibinfo {pages} {153004} (\bibinfo {year}
  {2008})}\BibitemShut {NoStop}%
\bibitem [{\citenamefont {Ullrich}(2006)}]{U06}%
  \BibitemOpen
  \bibfield  {author} {\bibinfo {author} {\bibfnamefont {C.~A.}\ \bibnamefont
  {Ullrich}},\ }\href {\doibase 10.1063/1.2406069} {\bibfield  {journal}
  {\bibinfo  {journal} {The Journal of Chemical Physics}\ }\textbf {\bibinfo
  {volume} {125}},\ \bibinfo {pages} {234108} (\bibinfo {year} {2006})},\
  \Eprint {http://arxiv.org/abs/https://doi.org/10.1063/1.2406069}
  {https://doi.org/10.1063/1.2406069} \BibitemShut {NoStop}%
\bibitem [{\citenamefont {Maitra}\ \emph
  {et~al.}(2002{\natexlab{a}})\citenamefont {Maitra}, \citenamefont {Burke},
  \citenamefont {Appel},\ and\ \citenamefont {Gross}}]{MBAG02}%
  \BibitemOpen
  \bibfield  {author} {\bibinfo {author} {\bibfnamefont {N.~T.}\ \bibnamefont
  {Maitra}}, \bibinfo {author} {\bibfnamefont {K.}~\bibnamefont {Burke}},
  \bibinfo {author} {\bibfnamefont {H.}~\bibnamefont {Appel}}, \ and\ \bibinfo
  {author} {\bibfnamefont {E.}~\bibnamefont {Gross}},\ }\href@noop {}
  {\bibfield  {journal} {\bibinfo  {journal} {Reviews of Modern Quantum
  Chemistry}\ }\textbf {\bibinfo {volume} {2}},\ \bibinfo {pages} {1186}
  (\bibinfo {year} {2002}{\natexlab{a}})}\BibitemShut {NoStop}%
\bibitem [{\citenamefont {Schaffhauser}\ and\ \citenamefont
  {K\"ummel}(2016)}]{SK16b}%
  \BibitemOpen
  \bibfield  {author} {\bibinfo {author} {\bibfnamefont {P.}~\bibnamefont
  {Schaffhauser}}\ and\ \bibinfo {author} {\bibfnamefont {S.}~\bibnamefont
  {K\"ummel}},\ }\href {\doibase 10.1103/PhysRevB.93.035115} {\bibfield
  {journal} {\bibinfo  {journal} {Phys. Rev. B}\ }\textbf {\bibinfo {volume}
  {93}},\ \bibinfo {pages} {035115} (\bibinfo {year} {2016})}\BibitemShut
  {NoStop}%
\bibitem [{\citenamefont {Thiele}\ and\ \citenamefont
  {K\"ummel}(2009)}]{TK09b}%
  \BibitemOpen
  \bibfield  {author} {\bibinfo {author} {\bibfnamefont {M.}~\bibnamefont
  {Thiele}}\ and\ \bibinfo {author} {\bibfnamefont {S.}~\bibnamefont
  {K\"ummel}},\ }\href {\doibase 10.1103/PhysRevA.79.052503} {\bibfield
  {journal} {\bibinfo  {journal} {Phys. Rev. A}\ }\textbf {\bibinfo {volume}
  {79}},\ \bibinfo {pages} {052503} (\bibinfo {year} {2009})}\BibitemShut
  {NoStop}%
\bibitem [{\citenamefont {Gross}\ and\ \citenamefont {Maitra}(2012)}]{GM12}%
  \BibitemOpen
  \bibfield  {author} {\bibinfo {author} {\bibfnamefont {E.}~\bibnamefont
  {Gross}}\ and\ \bibinfo {author} {\bibfnamefont {N.}~\bibnamefont {Maitra}},\
  }in\ \href@noop {} {\emph {\bibinfo {booktitle} {Fundamentals of
  Time-Dependent Density Functional Theory}}},\ \bibinfo {editor} {edited by\
  \bibinfo {editor} {\bibfnamefont {M.~A.}\ \bibnamefont {Marques}}, \bibinfo
  {editor} {\bibfnamefont {N.~T.}\ \bibnamefont {Maitra}}, \bibinfo {editor}
  {\bibfnamefont {F.~M.}\ \bibnamefont {Nogueira}}, \bibinfo {editor}
  {\bibfnamefont {E.}~\bibnamefont {Gross}}, \ and\ \bibinfo {editor}
  {\bibfnamefont {A.}~\bibnamefont {Rubio}}}\ (\bibinfo  {publisher} {Springer
  Berlin Heidelberg},\ \bibinfo {year} {2012})\ pp.\ \bibinfo {pages}
  {53--99}\BibitemShut {NoStop}%
\bibitem [{\citenamefont {Andrade}\ \emph {et~al.}(2018)\citenamefont
  {Andrade}, \citenamefont {Hamel},\ and\ \citenamefont {Correa}}]{AHC18}%
  \BibitemOpen
  \bibfield  {author} {\bibinfo {author} {\bibfnamefont {X.}~\bibnamefont
  {Andrade}}, \bibinfo {author} {\bibfnamefont {S.}~\bibnamefont {Hamel}}, \
  and\ \bibinfo {author} {\bibfnamefont {A.}~\bibnamefont {Correa}},\
  }\href@noop {} {\bibfield  {journal} {\bibinfo  {journal} {Eur. Phys. J. B.}\
  }\textbf {\bibinfo {volume} {91}},\ \bibinfo {pages} {229} (\bibinfo {year}
  {2018})}\BibitemShut {NoStop}%
\bibitem [{\citenamefont {Takeuchi}\ \emph {et~al.}(2019)\citenamefont
  {Takeuchi}, \citenamefont {Noda},\ and\ \citenamefont {Yabana}}]{TNY19}%
  \BibitemOpen
  \bibfield  {author} {\bibinfo {author} {\bibfnamefont {T.}~\bibnamefont
  {Takeuchi}}, \bibinfo {author} {\bibfnamefont {M.}~\bibnamefont {Noda}}, \
  and\ \bibinfo {author} {\bibfnamefont {K.}~\bibnamefont {Yabana}},\ }\href
  {\doibase 10.1021/acsphotonics.9b00889} {\bibfield  {journal} {\bibinfo
  {journal} {ACS Photonics}\ }\textbf {\bibinfo {volume} {6}},\ \bibinfo
  {pages} {2517} (\bibinfo {year} {2019})},\ \Eprint
  {http://arxiv.org/abs/https://doi.org/10.1021/acsphotonics.9b00889}
  {https://doi.org/10.1021/acsphotonics.9b00889} \BibitemShut {NoStop}%
\bibitem [{\citenamefont {Feist}(2009)}]{Feistthesis}%
  \BibitemOpen
  \bibfield  {author} {\bibinfo {author} {\bibfnamefont {J.}~\bibnamefont
  {Feist}},\ }\href@noop {} {\emph {\bibinfo {title} {Two-photon double
  ionization of helium, PhD Thesis}}}\ (\bibinfo {year} {2009})\BibitemShut
  {NoStop}%
\bibitem [{\citenamefont {Feist}\ \emph {et~al.}(2008)\citenamefont {Feist},
  \citenamefont {Nagele}, \citenamefont {Pazourek}, \citenamefont {Persson},
  \citenamefont {Schneider}, \citenamefont {Collins},\ and\ \citenamefont
  {Burgd\"orfer}}]{Feist08}%
  \BibitemOpen
  \bibfield  {author} {\bibinfo {author} {\bibfnamefont {J.}~\bibnamefont
  {Feist}}, \bibinfo {author} {\bibfnamefont {S.}~\bibnamefont {Nagele}},
  \bibinfo {author} {\bibfnamefont {R.}~\bibnamefont {Pazourek}}, \bibinfo
  {author} {\bibfnamefont {E.}~\bibnamefont {Persson}}, \bibinfo {author}
  {\bibfnamefont {B.~I.}\ \bibnamefont {Schneider}}, \bibinfo {author}
  {\bibfnamefont {L.~A.}\ \bibnamefont {Collins}}, \ and\ \bibinfo {author}
  {\bibfnamefont {J.}~\bibnamefont {Burgd\"orfer}},\ }\href {\doibase
  10.1103/PhysRevA.77.043420} {\bibfield  {journal} {\bibinfo  {journal} {Phys.
  Rev. A}\ }\textbf {\bibinfo {volume} {77}},\ \bibinfo {pages} {043420}
  (\bibinfo {year} {2008})}\BibitemShut {NoStop}%
\bibitem [{\citenamefont {Maitra}\ \emph
  {et~al.}(2002{\natexlab{b}})\citenamefont {Maitra}, \citenamefont {Burke},\
  and\ \citenamefont {Woodward}}]{MBW02}%
  \BibitemOpen
  \bibfield  {author} {\bibinfo {author} {\bibfnamefont {N.~T.}\ \bibnamefont
  {Maitra}}, \bibinfo {author} {\bibfnamefont {K.}~\bibnamefont {Burke}}, \
  and\ \bibinfo {author} {\bibfnamefont {C.}~\bibnamefont {Woodward}},\ }\href
  {\doibase 10.1103/PhysRevLett.89.023002} {\bibfield  {journal} {\bibinfo
  {journal} {Phys. Rev. Lett.}\ }\textbf {\bibinfo {volume} {89}},\ \bibinfo
  {pages} {023002} (\bibinfo {year} {2002}{\natexlab{b}})}\BibitemShut
  {NoStop}%
\bibitem [{\citenamefont {van Leeuwen}(1999)}]{L99}%
  \BibitemOpen
  \bibfield  {author} {\bibinfo {author} {\bibfnamefont {R.}~\bibnamefont {van
  Leeuwen}},\ }\href@noop {} {\bibfield  {journal} {\bibinfo  {journal} {Phys.
  Rev. Lett.}\ }\textbf {\bibinfo {volume} {82}},\ \bibinfo {pages} {3863}
  (\bibinfo {year} {1999})}\BibitemShut {NoStop}%
\bibitem [{\citenamefont {Ruggenthaler}\ and\ \citenamefont {van
  Leeuwen~R.}(2011)}]{RL11}%
  \BibitemOpen
  \bibfield  {author} {\bibinfo {author} {\bibfnamefont {M.}~\bibnamefont
  {Ruggenthaler}}\ and\ \bibinfo {author} {\bibnamefont {van Leeuwen~R.}},\
  }\href@noop {} {\bibfield  {journal} {\bibinfo  {journal} {Europhys. Lett.}\
  }\textbf {\bibinfo {volume} {95}},\ \bibinfo {pages} {13001} (\bibinfo {year}
  {2011})}\BibitemShut {NoStop}%
\bibitem [{\citenamefont {Nielsen}\ \emph {et~al.}(2013)\citenamefont
  {Nielsen}, \citenamefont {Ruggenthaler},\ and\ \citenamefont {van
  Leeuwen}}]{NRL13}%
  \BibitemOpen
  \bibfield  {author} {\bibinfo {author} {\bibfnamefont {S.~E.~B.}\
  \bibnamefont {Nielsen}}, \bibinfo {author} {\bibfnamefont {M.}~\bibnamefont
  {Ruggenthaler}}, \ and\ \bibinfo {author} {\bibfnamefont {R.}~\bibnamefont
  {van Leeuwen}},\ }\href {http://stacks.iop.org/0295-5075/101/i=3/a=33001}
  {\bibfield  {journal} {\bibinfo  {journal} {EPL (Europhysics Letters)}\
  }\textbf {\bibinfo {volume} {101}},\ \bibinfo {pages} {33001} (\bibinfo
  {year} {2013})}\BibitemShut {NoStop}%
\bibitem [{\citenamefont {Ruggenthaler}\ \emph {et~al.}(2015)\citenamefont
  {Ruggenthaler}, \citenamefont {Penz},\ and\ \citenamefont {van
  Leeuwen}}]{RPL15}%
  \BibitemOpen
  \bibfield  {author} {\bibinfo {author} {\bibfnamefont {M.}~\bibnamefont
  {Ruggenthaler}}, \bibinfo {author} {\bibfnamefont {M.}~\bibnamefont {Penz}},
  \ and\ \bibinfo {author} {\bibfnamefont {R.}~\bibnamefont {van Leeuwen}},\
  }\href@noop {} {\bibfield  {journal} {\bibinfo  {journal} {J. Phys. Condens.
  Matter}\ }\textbf {\bibinfo {volume} {27}},\ \bibinfo {pages} {203202}
  (\bibinfo {year} {2015})}\BibitemShut {NoStop}%
\bibitem [{\citenamefont {Bruner}\ \emph {et~al.}(2017)\citenamefont {Bruner},
  \citenamefont {Hernandez}, \citenamefont {Mauger}, \citenamefont {Abanador},
  \citenamefont {LaMaster}, \citenamefont {Gaarde}, \citenamefont {Schafer},\
  and\ \citenamefont {Lopata}}]{BHMALGSL18}%
  \BibitemOpen
  \bibfield  {author} {\bibinfo {author} {\bibfnamefont {A.}~\bibnamefont
  {Bruner}}, \bibinfo {author} {\bibfnamefont {S.}~\bibnamefont {Hernandez}},
  \bibinfo {author} {\bibfnamefont {F.}~\bibnamefont {Mauger}}, \bibinfo
  {author} {\bibfnamefont {P.~M.}\ \bibnamefont {Abanador}}, \bibinfo {author}
  {\bibfnamefont {D.~J.}\ \bibnamefont {LaMaster}}, \bibinfo {author}
  {\bibfnamefont {M.~B.}\ \bibnamefont {Gaarde}}, \bibinfo {author}
  {\bibfnamefont {K.~J.}\ \bibnamefont {Schafer}}, \ and\ \bibinfo {author}
  {\bibfnamefont {K.}~\bibnamefont {Lopata}},\ }\href {\doibase
  10.1021/acs.jpclett.7b01652} {\bibfield  {journal} {\bibinfo  {journal} {The
  Journal of Physical Chemistry Letters}\ }\textbf {\bibinfo {volume} {8}},\
  \bibinfo {pages} {3991} (\bibinfo {year} {2017})}\BibitemShut {NoStop}%
\bibitem [{\citenamefont {Chen}\ and\ \citenamefont {Lopata}(2020)}]{CL20}%
  \BibitemOpen
  \bibfield  {author} {\bibinfo {author} {\bibfnamefont {M.}~\bibnamefont
  {Chen}}\ and\ \bibinfo {author} {\bibfnamefont {K.}~\bibnamefont {Lopata}},\
  }\href {\doibase 10.1021/acs.jctc.0c00122} {\bibfield  {journal} {\bibinfo
  {journal} {Journal of Chemical Theory and Computation}\ }\textbf {\bibinfo
  {volume} {16}},\ \bibinfo {pages} {4470} (\bibinfo {year} {2020})},\ \bibinfo
  {note} {pMID: 32470295},\ \Eprint
  {http://arxiv.org/abs/https://doi.org/10.1021/acs.jctc.0c00122}
  {https://doi.org/10.1021/acs.jctc.0c00122} \BibitemShut {NoStop}%
\bibitem [{Note1()}]{Note1}%
  \BibitemOpen
  \bibinfo {note} {It is not {\protect \it a priori} obvious that the unique
  solution to Eq.~(\ref {eq:continuity}) with the boundary-condition Eq.~(\ref
  {boundary conditions}) applied with time $t$ as a parameter is compatible
  with the TDKS evolution, but as the results evolve smoothly in time, it
  is.}\BibitemShut {Stop}%
\bibitem [{\citenamefont {Brus}(2014)}]{Brus2014}%
  \BibitemOpen
  \bibfield  {author} {\bibinfo {author} {\bibfnamefont {L.}~\bibnamefont
  {Brus}},\ }\href {\doibase 10.1021/ar500175h} {\bibfield  {journal} {\bibinfo
   {journal} {Accounts of Chemical Research}\ }\textbf {\bibinfo {volume}
  {47}},\ \bibinfo {pages} {2951} (\bibinfo {year} {2014})},\ \bibinfo {note}
  {pMID: 25120173},\ \Eprint
  {http://arxiv.org/abs/https://doi.org/10.1021/ar500175h}
  {https://doi.org/10.1021/ar500175h} \BibitemShut {NoStop}%
\bibitem [{\citenamefont {Kraisler}\ \emph {et~al.}(2021)\citenamefont
  {Kraisler}, \citenamefont {Hodgson},\ and\ \citenamefont {Gross}}]{KHG21}%
  \BibitemOpen
  \bibfield  {author} {\bibinfo {author} {\bibfnamefont {E.}~\bibnamefont
  {Kraisler}}, \bibinfo {author} {\bibfnamefont {M.~J.~P.}\ \bibnamefont
  {Hodgson}}, \ and\ \bibinfo {author} {\bibfnamefont {E.~K.~U.}\ \bibnamefont
  {Gross}},\ }\href {\doibase 10.1021/acs.jctc.0c01093} {\bibfield  {journal}
  {\bibinfo  {journal} {Journal of Chemical Theory and Computation}\ }\textbf
  {\bibinfo {volume} {17}},\ \bibinfo {pages} {1390} (\bibinfo {year}
  {2021})},\ \bibinfo {note} {pMID: 33595312},\ \Eprint
  {http://arxiv.org/abs/https://doi.org/10.1021/acs.jctc.0c01093}
  {https://doi.org/10.1021/acs.jctc.0c01093} \BibitemShut {NoStop}%
\bibitem [{\citenamefont {Maitra}(2017)}]{M17}%
  \BibitemOpen
  \bibfield  {author} {\bibinfo {author} {\bibfnamefont {N.~T.}\ \bibnamefont
  {Maitra}},\ }\href {http://stacks.iop.org/0953-8984/29/i=42/a=423001}
  {\bibfield  {journal} {\bibinfo  {journal} {Journal of Physics: Condensed
  Matter}\ }\textbf {\bibinfo {volume} {29}},\ \bibinfo {pages} {423001}
  (\bibinfo {year} {2017})}\BibitemShut {NoStop}%
\bibitem [{\citenamefont {van Gisbergen}\ \emph {et~al.}(1999)\citenamefont
  {van Gisbergen}, \citenamefont {Schipper}, \citenamefont {Gritsenko},
  \citenamefont {Baerends}, \citenamefont {Snijders}, \citenamefont
  {Champagne},\ and\ \citenamefont {Kirtman}}]{GSGB99}%
  \BibitemOpen
  \bibfield  {author} {\bibinfo {author} {\bibfnamefont {S.~J.~A.}\
  \bibnamefont {van Gisbergen}}, \bibinfo {author} {\bibfnamefont {P.~R.~T.}\
  \bibnamefont {Schipper}}, \bibinfo {author} {\bibfnamefont {O.~V.}\
  \bibnamefont {Gritsenko}}, \bibinfo {author} {\bibfnamefont {E.~J.}\
  \bibnamefont {Baerends}}, \bibinfo {author} {\bibfnamefont {J.~G.}\
  \bibnamefont {Snijders}}, \bibinfo {author} {\bibfnamefont {B.}~\bibnamefont
  {Champagne}}, \ and\ \bibinfo {author} {\bibfnamefont {B.}~\bibnamefont
  {Kirtman}},\ }\href {\doibase 10.1103/PhysRevLett.83.694} {\bibfield
  {journal} {\bibinfo  {journal} {Phys. Rev. Lett.}\ }\textbf {\bibinfo
  {volume} {83}},\ \bibinfo {pages} {694} (\bibinfo {year} {1999})}\BibitemShut
  {NoStop}%
\bibitem [{\citenamefont {Gori-Giorgi}\ \emph {et~al.}(2016)\citenamefont
  {Gori-Giorgi}, \citenamefont {G\'al},\ and\ \citenamefont
  {Baerends}}]{GGB16}%
  \BibitemOpen
  \bibfield  {author} {\bibinfo {author} {\bibfnamefont {P.}~\bibnamefont
  {Gori-Giorgi}}, \bibinfo {author} {\bibfnamefont {T.}~\bibnamefont {G\'al}},
  \ and\ \bibinfo {author} {\bibfnamefont {E.~J.}\ \bibnamefont {Baerends}},\
  }\href {\doibase 10.1080/00268976.2015.1137643} {\bibfield  {journal}
  {\bibinfo  {journal} {Molecular Physics}\ }\textbf {\bibinfo {volume}
  {114}},\ \bibinfo {pages} {1086} (\bibinfo {year} {2016})},\ \Eprint
  {http://arxiv.org/abs/https://doi.org/10.1080/00268976.2015.1137643}
  {https://doi.org/10.1080/00268976.2015.1137643} \BibitemShut {NoStop}%
\bibitem [{\citenamefont {Umrigar}\ and\ \citenamefont {Gonze}(1994)}]{UG94}%
  \BibitemOpen
  \bibfield  {author} {\bibinfo {author} {\bibfnamefont {C.~J.}\ \bibnamefont
  {Umrigar}}\ and\ \bibinfo {author} {\bibfnamefont {X.}~\bibnamefont
  {Gonze}},\ }\href {\doibase 10.1103/PhysRevA.50.3827} {\bibfield  {journal}
  {\bibinfo  {journal} {Phys. Rev. A}\ }\textbf {\bibinfo {volume} {50}},\
  \bibinfo {pages} {3827} (\bibinfo {year} {1994})}\BibitemShut {NoStop}%
\end{thebibliography}%

\end{document}